# Assessment of Cr isotopic heterogeneities of volatile-rich asteroids based on multiple planet formation models

Ryota FUKAI*[1,2] and Sota ARAKAWA[3,4]

1: Geochemical Research Center, Graduate School of Science, The University of Tokyo

2: Institute of Space and Astronautical Science, Japan Aerospace Exploration Agency

3: Department of Earth and Planetary Sciences, School of Science, Tokyo Institute of Technology

4: Division of Science, National Astronomical Observatory of Japan

* Corresponding author (ryota.fukai@gmail.com)

# Abstract

Describing the comprehensive evolutionary scenario for asteroids is key to explaining the various physical processes of the solar system. Bulk-scale carbonaceous chondrites (CCs) possibly record the primordial information associated with the formation processes of their parent bodies. In this study, we tried to estimate the relative formation region of volatile-rich asteroids by utilizing the nucleosynthetic Cr isotopic variation ($^{54}$Cr/$^{52}$Cr) in bulk-scale CCs. Numerical calculations were conducted to track the temporal evolution of isotopically different (solar and presolar) dust and $^{54}$Cr/$^{52}$Cr values for mixed materials with disk radius. First, we found that isotopic heterogeneities in CC formation regions would be preserved with a weak turbulence setting that would increase the timescales of the advection and diffusion in the disk. Second, we assessed the effects of gaps formed by giant planets. Finally, the distance from the injected supernovae and Cr isotopic compositions of the presolar grains were investigated in terms of the estimated formation region of CCs. In our results, a plausible formation region of four types of CCs can be obtained with the supernova from approximately 2 pc and typical Cr isotopic compositions of presolar grains. Among the parent bodies of CCs (i.e., volatile-rich asteroids), B-type asteroids formed in the outermost region, which is inconsistent with the present population, showing that D-type asteroids are generally located beyond most of the C-complex asteroids. Both the initial and present orbits of asteroids might be explained by the scatter attributed to the inward-outward migration of Jupiter and Saturn.

# 1. Introduction

The C-complex asteroids located in the main asteroid belt are considered to have formed in lower-temperature environments compared with the formation region of rocky planets. These types of small bodies should possess the vital records of several types of volatiles, such as ice and organics, in the early Solar System. To explain the chemical compositions of bulk silicate earth, these volatiles were possibly delivered to the rocky planets in the early stage of their formation, or by late accretion after their formation (e.g., Morbidelli et al. 2000). Thus, deciphering the formation processes of C-complex asteroids is essential not only for establishing a comprehensive evolutionary scenario of the asteroid belt but also for constraining the origin of volatiles in rocky planets. However, spatial information associated with the asteroid formation process in the protoplanetary disk has not been clearly identified because the initial orbits of asteroids were likely disturbed by interactions with gas, other small bodies, and planets.

The original information on the formation region of C-complex asteroids was recorded in carbonaceous chondrites (CCs). CCs are undifferentiated meteorites considered to originate from C-complex (partly D-, K-, and X-type) asteroids based on the similarity in reflectance spectroscopy (e.g., Cloutis et al. 2011a; 2011b; 2012; Vernazza et al. 2015; 2017). Cosmochemical studies revealed that CCs show variable isotopic compositions because of the nucleosynthetic origin in the Fe-peak (Warren 2011) and trans-Fe (Fukai et al. 2017; 2019; Kruijer et al. 2019; Yokoyama et al. 2019) elements. The isotopic compositions of bulk CCs vary with the spatial and/or temporal heterogeneities of "presolar grains" that possess extremely anomalous isotopic compositions compared to bulk CCs (Dauphas & Schauble 2016). In particular, Cr isotopic compositions ($\varepsilon^{54}$Cr) discriminate well among the eight types of CCs (Trinquier et al. 2007) possibly induced by the incorporation of highly anomalous Cr oxide grains (Dauphas et al. 2010; Qin et al. 2011; Nittler et al. 2018).

The physicochemical nebula processes that induce nucleosynthetic isotopic variation have been extensively discussed in several previous studies. One plausible model of Cr isotopic heterogeneity was proposed to have occurred within the initially homogeneous solar nebula. The depletion of moderately volatile elements in CCs (such as Ge) seems to correlate with the degree of Cr isotopic anomalies, which led to the conclusion that isotopic heterogeneity was achieved with selective thermal destruction of presolar grains (Trinquier et al. 2009). However, the feasibility of the processes in the CC formation region has not been certified in terms of the refractory characteristics of Cr ($T_c$, 50% condensation temperature: 1296 K; Lodders 2003).

Alternatively, infall from inhomogeneous molecular cloud core or supernova injection of presolar dust into the solar nebula may explain the observed isotopic variation. A two-stage infall from an isotopically heterogeneous molecular cloud core was invoked to explain the Cr isotopic variation within calcium-aluminum-rich inclusions (CAIs), non-carbonaceous meteorites (NCs), and CCs  (Jacquet et al. 2019; Nanne et al. 2019). However, the infall model may possess two difficulties: that the mass and compositions of the infallen materials are difficult to estimate and that the origin of the isotopically inhomogeneous molecular cloud core has not been explained yet. In contrast, a previous study explained the Cr isotopic variation among bulk-scale meteorites by nearby supernova injection (Sugiura & Fujiya 2014). They considered the inward diffusion of presolar materials injected at a large distance from the solar nebula and suggested that Cr isotopic heterogeneity was achieved with temporal changes in the nebula. However, recent studies revealed that NCs and CCs show the fundamental isotopic dichotomy of Ti–Cr (Warren 2011) and Mo isotopes (Kruijer et al. 2019). This observation leads to the conclusion that the isotopic heterogeneities were accomplished due to spatial reasons, mainly with the separation of NCs and CCs by an early-formed Jupiter.

In this study, we reassessed the physical condition of the protoplanetary disk to reproduce the $\varepsilon^{54}$Cr variations in numerical calculations by introducing the supernova injection model (Ouellette et al. 2010) and gaps formed by early formed giant planets (Kruijer et al. 2017; Desch et al. 2018). After investigating the effects of advection, diffusion, drift, and gaps in the disk model, we show that Cr isotopic heterogeneities of the solar nebula can be explained by uniform supernova injection with relatively low turbulence intensity. We further suspect the relative formation regions of two types of C-complex asteroids (i.e., B and Ch types) and two other types of volatile-rich asteroids (i.e., D and K types) by using the $\varepsilon^{54}$Cr profile in the CC formation region. Finally, we consider radial mixing in asteroid belts using the spatial information obtained from the asteroids.

# 2. Method

*2.1 Disk model with gas and dust evolution*

*2.1.1 Gas disk*

We calculated the viscous evolution of the protoplanetary disk and obtained the radial distribution of gas and dust particles as a function of time, similar to that obtained in earlier studies (Yang & Ciesla 2012; Desch et al. 2018). The initial distribution of the gas surface density, $\Sigma_{\mathrm{gas,0}}$, is given by a self-similar profile, described as:

$$\Sigma_{\mathrm{gas,0}} = \frac{(2-\gamma)M_{\mathrm{d}}}{2\pi R_{\mathrm{d}}{}^2} \times \left(\frac{r}{R_{\mathrm{d}}}\right)^{-\gamma} \times \exp\left(-\left(\frac{r}{R_{\mathrm{d}}}\right)^{2-\gamma}\right), \quad (1)$$

where $r$ is the radial distance from the Sun (Hartmann et al. 1998). We set the disk exponent to $\gamma = 15/14$, initial disk radius to $R_{\mathrm{d}} = 5$ au, and total disk mass to $M_{\mathrm{d}} = 0.01\ M_{\mathrm{Sun}}$ (solar mass). The disk exponent of $\gamma = 15/14$ is usually chosen for passively heated disks, and the surface density structure is very close to that of the minimum mass solar nebula model (Hayashi 1981) when we set $R_{\mathrm{d}} = 5$ au and $M_{\mathrm{d}} = 0.01\ M_{\mathrm{Sun}}$. We then calculated the time evolution of the gas surface density, $\Sigma_{\mathrm{gas}}$. The basic equation of an accretion disk is:

$$\frac{\partial \Sigma_{\mathrm{gas}}}{\partial t} = \frac{1}{2\pi r}\frac{\partial \dot{M}_{\mathrm{gas}}}{\partial r}, \quad (2)$$

where

$$\dot{M}_{\mathrm{gas}} = 6\pi r^{1/2}\frac{\partial}{\partial r}\left(r^{1/2}\,\Sigma_{\mathrm{gas}}v_{\mathrm{acc}}\right) \quad (3)$$

is the mass flux at every location (Desch et al. 2018). The mass flux is proportional to the kinematic viscosity of the disk, $v_{\mathrm{acc}}$, which is given by:

$$v_{\mathrm{acc}} = \alpha_{\mathrm{acc}}c_s h_{\mathrm{g}}, \quad (4)$$

where $\alpha_{\mathrm{acc}}$ is the angular momentum transport efficiency parameter (Shakura & Sunyaev

1973), $c_s$ is the sound speed, and $h_g$ is the gas scale height. The gas scale height is given by $h_g = c_s/\Omega_K$, where $\Omega_K$ is the Keplerian frequency. The timescale of the viscous evolution of the gas disk is inversely proportional to $\nu_{acc}$; hence, $\alpha_{acc}$ is the key parameter for understanding the time evolution of the protoplanetary disk (Ciesla 2010; Sugiura & Fujiya 2014). The temperature profile adopted herein is a passively heated disk model (Kusaka et al. 1970; Chiang & Goldreich 1997) as follows: $T = 163 \ (r \ / \ 1 \ \text{au})^{-3/7}$, where $T$ is the disk temperature.

We also considered the effects of the formation of planetary cores of giant planets (i.e., Jupiter and Saturn) on disk evolution. We artificially increased $\alpha_{acc}$ in the vicinity of the giant planets to mimic the planetary gap in the disk (Desch, et al. 2018). The effect of the planetary gap is included in our numerical code as follows:

$$\alpha_{acc} = \alpha \left[ 1 + A \exp\left( - \left[ \frac{(r - r_p)}{B} \right]^2 \right) \right], \ (5)$$

where $\alpha$ is the angular momentum transport efficiency parameter outside the gap, $r_p$ is the location of the planetary core, and $A$ and $B$ are the parameters that control the geometry of the gas gap. We calculated the depth and width of the gas gap formed by the planet using the empirical formulae (Kanagawa et al. 2015a; 2015b; 2016). The dimensionless parameter $K$ introduced by Kanagawa et al. (2015b) is related to the gap depth, and $K$ is given by:

$$K = \left( \frac{M_p}{M_{Sun}} \right)^2 \left( \frac{h_g}{r_p} \right)^{-5} \alpha^{-1}, \quad (6)$$

where $M_p$ is the mass of the planetary core. The gas surface density at the bottom of the gap is proportional to $(1 + 0.04 \ K)^{-1}$. The full width of the gap measured at the level of half of the surface density outside the gap region, $\Delta$, is also given by Kanagawa et al. (2016) as follows:

$$\frac{\Delta}{r_p} = 0.41 \left( \frac{M_p}{M_{Sun}} \right)^{1/2} \left( \frac{h_g}{r_p} \right)^{-3/4} \alpha^{-1/4}. \quad (7)$$

As the gas surface density of a passively heated accretion disk is inversely proportional to $\alpha_{acc}$, we can rewrite Eq. (5), as follows:

$$\alpha_{acc} = \alpha \left[ 1 + 0.04\, K \exp\left( - \left[ \frac{(r - r_p)}{\Delta/2} \right]^2 \times \max[1,\ \log(0.04\, K)] \right) \right]. \quad (8)$$

We assumed multiple settings regarding the formation region and timing of giant planets. The formation regions of Jupiter and Saturn are based on terrestrial planet-forming models (Walsh et al. 2011). We also follow the suggestion of the early formation (0.5–2.0 Myr) of the core of Jupiter by previous studies (Kruijer et al. 2019; Mousis et al. 2019).

*2.1.2 Dust particles*

The following two dust particle populations were introduced: solar and presolar components. The initial surface density of the solar dust $\Sigma_{SD,0}$ (= 0.01 $\Sigma_{gas,0}$) decreased with increasing $r$. In contrast, we assumed that the presolar component was injected from a nearby supernova at the initial stage of disk evolution (Ouellette et al. 2007). In this case, the surface density of the injected presolar dust is approximately constant within the disk radius (see Ouellette et al. 2009; 2010) and the cutoff radius may be close to the disk radius. We then set the surface density of the presolar dust, $\Sigma_{PSD,0}$, which is described as:

$$\Sigma_{PSD,0} = \Sigma_{inj} \times \exp\left( -\left( \frac{r}{R_d} \right)^{2-\gamma} \right). \quad (9)$$

We parameterized $\Sigma_{inj}$ in this study. We then calculated the temporal evolution of the dust surface densities (solar and presolar) considering three physical processes, namely advection, diffusion, and the drift of dust particles (see Desch et al. 2017; 2018).

The surface densities of solar and presolar dust components $\Sigma_{SD}$ and $\Sigma_{PSD}$ evolve according to:

$$\frac{\partial \Sigma_i}{\partial t} = \frac{1}{2\pi r} \frac{\partial \dot{M}_i}{\partial r} \ (i = SD\ or\ PSD), \quad (10)$$

and the mass flux of dust particles is given by the sum of three processes:

$$\dot{M}_i = \dot{M}_{i,\text{adv}} + \dot{M}_{i,\text{diff}} + \dot{M}_{i,\text{drift}}. \tag{11}$$

The effect of advection is given by:

$$\dot{M}_{i,\text{adv}} = \frac{\Sigma_i}{\Sigma_{\text{gas}}} \dot{M}_{\text{gas}} = 2\pi r \Sigma_i v_{\text{adv}}. \tag{12}$$

The effect of diffusion is introduced by adding a term:

$$\dot{M}_{i,\text{diff}} = 2\pi r \Sigma_{\text{gas}} D \frac{\partial}{\partial r}\left(\frac{\Sigma_i}{\Sigma_{\text{gas}}}\right) = 2\pi r \Sigma_i v_{\text{diff}}, \tag{13}$$

and

$$D = \frac{\nu}{\text{Sc}(1+\text{St}^2)} \tag{14}$$

is the diffusion coefficient of the dust particles. We assume a Schmidt number of $\text{Sc} = 0.7$ (Desch et al. 2017; 2018). We consider two cases for the treatment of viscosity $\nu = \alpha_{\text{diff}} c_s h_g$: (i) $\alpha_{\text{diff}} = \alpha_{\text{acc}}$ as Desch et al. (2018) assumed for simplicity and (ii) $\alpha_{\text{diff}} = \alpha \neq \alpha_{\text{acc}}$, which might be more plausible (see Sec. 3.2). The Stokes number St is given by:

$$\text{St} = \Omega_K \frac{\rho_{\text{agg}} a}{\rho_g \sqrt{8/\pi} c_s}, \tag{15}$$

where $\rho_g = \Sigma_{\text{gas}}/\left(\sqrt{2\pi} h_g\right)$ is the midplane gas density, $\rho_{\text{agg}}$ the density of dust aggregates, and $a$ is the aggregate radius. The radial drift of dust particles also affects the temporal evolution of the dust surface density. This effect is given by

$$\dot{M}_{i,\text{drift}} = 2\pi r \Sigma_i \Delta u, \tag{16}$$

and $\Delta u$ is the radial drift velocity of the dust particles with respect to the gas. The radial drift velocity also depends on the Stokes number as follows:

$$\Delta u = \frac{\mathrm{St}}{1+\mathrm{St}^2}(\eta r \Omega_{\mathrm{K}} - \mathrm{St} v_{\mathrm{adv}}), \qquad (17)$$

and the normalized pressure gradient $\eta$ is given by

$$\eta = -\frac{1}{r\Omega_{\mathrm{K}}^2}\frac{1}{\rho_{\mathrm{g}}}\frac{\partial}{\partial r}\left(\rho_{\mathrm{g}} c_{\mathrm{s}}^2\right). \qquad (18)$$

Equation (15) shows that St depends on $\rho_{\mathrm{agg}}a$ and we introduce "the equivalent radius" $a_{\mathrm{eq}}$ as follows:

$$a_{\mathrm{eq}} = a\frac{\rho_{\mathrm{agg}}}{3\,\mathrm{g\,cm}^{-3}}. \qquad (19)$$

The particle radius of the solar and presolar dust was set to the grains observed in meteorites without any size distribution and dust growth ($a_{\mathrm{eq}} = 10$ nm, 100 nm, 1 μm, and 10 μm). Presolar grains have a size distribution from approximately 0.01 μm to approximately 10 μm (Amari et al. 1994) and maximum peak in the range of 100 to 200 nm in presolar silicates (Hoppe et al. 2017), although most of the grains may be immediately coagulated with the other grains and evolve into similar-sized dust aggregates (e.g., Okuzumi et al. 2012). Collisions between solar and presolar components also occur frequently, and the assumption that solar and presolar dust aggregates have similar particle sizes may work well.

*2.2 Cr isotope evolutions*

The Cr isotope anomalies are described in the ε notation, which represents parts per $10^4$ deviations from the mean values of the terrestrial reference material ($\varepsilon^{54}\mathrm{Cr} = [(^{54}\mathrm{Cr}/^{52}\mathrm{Cr}_{\mathrm{meteorites}}) / (^{54}\mathrm{Cr}/^{52}\mathrm{Cr}_{\mathrm{terrestrial}}) - 1] \times 10^4$). The $\varepsilon^{54}\mathrm{Cr}$ values with the disk radius ($r$) were calculated from the following mass balance equation:

$$\varepsilon^{54}Cr = \frac{(\varepsilon^{54}Cr)_{SD} \times \Sigma_{SD} + (\varepsilon^{54}Cr)_{PSD} \times \Sigma_{PSD}}{\Sigma_{SD} + \Sigma_{PSD}}. \qquad (20)$$

We assumed $(\varepsilon^{54}Cr)_{SD} = -1$ based on the lowest values among bulk-scale meteorites. This assumption was certified by the mass-balance calculation in a previous study (Sugiura & Fujiya 2014). We assumed $(\varepsilon^{54}Cr)_{PSD} = 25,000$ or $500,000$ from the measurement of the presolar oxide grains (Dauphas et al. 2010; Nittler et al. 2018). These grains are highly enriched in neutron-rich isotopes likely formed in the ejecta of supernovae. We also assumed that solar and presolar dust possess the same elemental abundances of Cr. This assumption was verified by mass-balance calculations to explain the bulk-scale Orgueil meteorite (Dauphas et al. 2010).

*2.3 Classification of meteorites and asteroids*

The eight subgroups of CCs have been linked to individual types of asteroids through the reflectance spectroscopy of the asteroid surfaces. The reflectance spectra for most CCs are generally matched to C-complex asteroids. C-complex asteroids are classified into five subtypes based on spectral information, such as albedo and absorption band (i.e., B, C, Cb, Ch, and Cgh types). CI chondrites are generally linked to B-type asteroids, which share a maximum reflectance of 0.55-μm bands (Cloutis et al. 2011a). CM chondrites have 3-μm-feature bands similar to those of Ch-type asteroids (Burbine 1998). In contrast, no good matches can be found between some types of CCs and C-complex asteroids. Tagish Lake, which is an ungrouped C2 meteorite, is generally linked to D-type asteroids mainly caused by highly red slopes (Hiroi et al. 2001). CV chondrites have shallow 1-μm bands that are typically linked to the Eos family (Greenwood et al. 2020) or other types of K-type asteroids (Burbine et al. 2001). In this study, the CCs and asteroids linked here are treated to share the Cr isotopic compositions throughout the evolution of the solar system.

The Cr isotopic anomalies of individual CCs were determined by high-precision isotope ratio measurements with a multi-collector-type mass spectrometer. Subgroups of CCs are well discriminated within each other based on the isotope data (Shukolyukov & Lugmair 2006; Trinquier et al. 2007; Qin et al. 2010; Larsen et al. 2011; Petitat et al. 2011; Jenniskens et al. 2012; Hibiya et al. 2018). Table 1 shows the data of the Cr isotopic anomalies for the meteorite groups discussed earlier (i.e., CI, CM, C2-ungrouped, and CV). We selected the most representative and repeatedly measured meteorites for the CI, CM, and CV groups as follows: Orgueil, Murchison, and Allende, respectively.

# 3. Results and Discussion

### 3.1 Distribution of solar and presolar dust

We first assessed a simple model without assuming any giant planets to demonstrate the general behavior of dust particles in the protoplanetary disk. We show the profile of $\varepsilon^{54}Cr$ values of mixed dust within 1 au to approximately 10 au for the turbulent parameters $\alpha = 10^{-3}$ and $= 10^{-4}$, which are typically used to track the evolutional model of the gas disk (e.g., Ciesla 2010). In this section, we fixed $\Sigma_{inj} = 2 \times 10^{-4}$ g cm$^{-2}$ and $(\varepsilon^{54}Cr)_{PSD} = 25,000$, which is a typical value constrained by the discussion in Sec. 3.3. In the initial states of the disk ($t = 0$ Myr), $\varepsilon^{54}Cr$ values increased from -1 to +5 with disk radius, which adequately reproduced the $\varepsilon^{54}Cr$ values obtained from solar system materials (dotted lines in Fig. 1). The outer system was enriched in the $^{54}Cr$ isotope because presolar dust was injected uniformly as compared with the initially settled solar dust. The enrichment of supernova-derived dust in the outer solar system was similarly observed in the calculation of Nanne et al. (2019). The $\varepsilon^{54}Cr$ values were completely homogenized to approximately +1 until $t = 4$ Myr in the case of $\alpha = 10^{-3}$ because the disk evolved rapidly (Fig. 1a–b). In contrast, the $\varepsilon^{54}Cr$ heterogeneity was preserved in the case of $\alpha = 10^{-4}$ (Fig. 1c–d). This consequence has already been inferred in Sugiura and Fujiya (2014); however, we introduced the advection and diffusion timescales for the disk ($\tau_{adv}, \tau_{diff}$) and radial drift ($\tau_{drift}$) for the dust particles to interpret the results more quantitatively.

Fig. 2 shows the timescales $\tau_{adv} \equiv |r/v_{adv}|$ and $\tau_{diff} \equiv |r/v_{diff}|$ with disk radius for each $\alpha$ and the equivalent radius of dust grains ($a_{eq}$). The advection timescale, $\tau_{adv}$, shows a gradual increase to the initial radius of the disk (5 au) and then a decrease to the outer edge of the disk. Both the advection and diffusion timescales, $\tau_{adv}$ and $\tau_{diff}$, were inversely proportional to $\alpha$ and were $< 10^6$ yr in the most range of a disk with $\alpha = 10^{-3}$ (Fig. 2a–b). This suggests that $\tau_{adv}$ and $\tau_{diff}$ dominantly control the speed of the disk evolution and the homogenization of $\varepsilon^{54}Cr$ values with $\alpha = 10^{-3}$.

However, $\tau_{adv}$ and $\tau_{diff}$ were $> 10^6$ yr, except for the most inner parts of the disk with $\alpha = 10^{-4}$ (Fig. 2c–d). As a result, the $\varepsilon^{54}Cr$ heterogeneity was almost fully preserved until $t = 4$ Myr with small and/or fluffy dust, $a_{eq} = 10^{-4}$ to $10^{-5}$ cm (Fig. 1d). However, even in $\alpha = 10^{-4}$, $\varepsilon^{54}Cr$ values were flattened, especially in the outer part of the disk with $a_{eq} = 10^{-2}$ cm to $10^{-3}$ cm (Fig. 1c). This suggests that the grain radius mainly controls how homogenize the disk with $\alpha = 10^{-4}$ is in contrast to the case of $\alpha = 10^{-3}$. In Fig. 2, we also show the timescale of the radial drift $\tau_{drift}$. For small grains whose St is smaller than 1, $\tau_{drift}$ was inversely correlated with $a_{eq}$. As $\tau_{drift}$ was $< 10^6$ yr within some part of the disk with $a_{eq} = 10^{-2}$ cm and $10^{-3}$ cm, dust was transported inward, the $\Sigma_{SD}$ and $\Sigma_{PSD}$ were completely depleted in the region (Fig. 3) and $\varepsilon^{54}$Cr seems to be homogenized due to the large gradient in the dust-to-gas ratio (see Eq. 13). We also showed that the Stokes number was less than unity through the simulation (Fig. 4), suggesting that $\tau_{diff}$ hardly depends on dust particle sizes when the radial drift of dust particles is negligible, that is, $a_{eq} = 10^{-4}$ cm and $10^{-5}$ cm. It should be noted that either the radial drift of dust particles occurred and the isotopic heterogeneity was not preserved at $\alpha = 10^{-3}$. In summary, weaker turbulence ($\alpha = 10^{-4}$) and small and/or fluffy dust population ($a_{eq} = 10^{-4}$–$10^{-5}$ cm) are suitable to avoid the rapid homogenization of the disk.

### 3.2 Effects of gap formed by giant planets

In general, early formed giant planets in the protoplanetary disk would deplete the materials belonging to their orbits and open a compositional gap. This would also lead to the concentration of larger dust at the local gas pressure maximum of the outer edge of the gap (Desch et al. 2018; Haugbølle et al. 2019). We assessed the effects of an open gap on $\varepsilon^{54}$Cr heterogeneities in detail by comparing the model discussed in Sec. 3.1. In this section, we fixed $\Sigma_{inj} = 2 \times 10^{-4}$ g cm$^{-2}$ and $(\varepsilon^{54}Cr)_{PSD} = 25,000$.

In this section, we explain the simulation with proto-Jupiter ($20M_E$) located in 3.5 au, followed by the concept of the Grand-Tack model (Walsh et al. 2011; Kruijer et al. 2017). It should be noted that the proto-Saturn ($20M_E$) formed in 4.5 au (Walsh et al. 2011) did not affect the profiles in $\varepsilon^{54}$Cr (see Appendix). We also tested the light proto-Jupiter case ($3M_E$) suggested by the low-mass asteroid belt model (Raymond & Izidoro 2017) but no critical changes were observed in the simulation without a Jupiter discussed in Sec. 3.1 (see Appendix).

We artificially increased $\alpha_{acc}$ in the vicinity of the giant planets to reproduce the effects of Jupiter (Sec. 2.1.1). First, we set $\alpha_{diff} = \alpha_{acc}$ following Desch et al. (2018). For $\alpha = 10^{-3}$, no changes from the simulation without giant planets were observed in the $\varepsilon^{54}$Cr profiles (Fig. 5a), although the surface densities of solar dust show a relative depletion of approximately 3.5 au (Fig. 6a). Critical changes were observed only in the case of $a_{eq} = 10^{-2}$ cm with $\alpha = 10^{-}$ [4] (Fig. 5b). In this case, the $\varepsilon^{54}$Cr slopes increased steeply around the vicinity of the giant planets. In addition, the surface densities of dust show a maximum peak around 4.0 au (Fig. 6b), which would lead to a drastic change in $\varepsilon^{54}$Cr values. However, $\varepsilon^{54}$Cr values were flattened beyond 4.0 au due to the depletion of dust. The isotopic heterogeneities observed in multiple types of CCs (Table 1) are difficult to explain with this $\varepsilon^{54}$Cr flatness.

We also assess the case of $\alpha_{diff}$ = const. (= $10^{-3}$ or $10^{-4}$). This assumption seems to be more plausible compared to the first assumption because angular momentum transfer in the vicinity of the planet is mainly caused by the gravitational torque exerted by a planet, not by the kinematic viscosity of the disk. At $\alpha_{diff}$ = $10^{-3}$, no changes were observed from the case of $\alpha_{diff}$ = $\alpha_{acc}$ because a small $\tau_{adv}$ (< $10^6$ yr) dominantly controls the homogenization of the disk (Fig. 7a, c). Clear steps in $\varepsilon^{54}$Cr around the vicinity of Jupiter were observed within $\alpha_{diff}$ = $10^{-4}$. The larger steps in this setting were possibly induced by the suppression of diffusion from the orbits beyond Jupiter.

In reality, meteoritical evidence suggests that $\varepsilon^{54}$Cr jumps between NCs and CCs reached approximately 0.8 (Warren 2011), which is consistent with the case of $\alpha$ = $10^{-4}$. From the conclusion in Sec. 3.1, the small and/or fluffy dust is suitable for reproducing the isotopic heterogeneities within CC reservoirs. Combined with these conclusions, low-$\alpha$ and small-$a_{eq}$ settings are the most suitable for reproducing the meteoritical $\varepsilon^{54}$Cr (Fig. 7d).

The grain size dependence discussed in Sec. 3.1–3.2 would also constrain the other carrier grains responsible for the isotopic dichotomy, including, for example, r-process enriched materials in the Mo isotopic space (Kruijer et al. 2017). The grain size distribution of silicon carbide grains condensed in supernovae (a candidate for r-process nuclide carrier grains) was $10^{-4}$–$10^{-5}$ cm (Amari et al. 1994; Kozasa et al. 1991), which is suitable for the preservation of the isotopic heterogeneity in the disk.

Also, our results suggest that some (unknown) processes that prevent μm-sized dust grains from growing into large and compact aggregates during the first a few million years of the solar nebula are needed to reproduce the meteoritical $\varepsilon^{54}$Cr (Okuzumi 2009). We note that this inefficient dust growth scenario is consistent with the spectral energy distributions of classical T Tauri stars (Dullemond & Dominik 2005) and observed the dust-to-gas mass ratio in the young protoplanetary disks (Homma & Nakamoto 2018).

*3.3 Injection and relative formation region of CCs*

As discussed in Sec. 3.1–3.2, turbulence and grain sizes are critical parameters for determining the degree of isotopic heterogeneities in the protoplanetary disk. In addition to these parameters, $\Sigma_{inj}$, representing the surface densities of the injected materials and $\varepsilon^{54}Cr$ values of the presolar materials are needed for evaluation. These values finalize the $\varepsilon^{54}Cr$ compositions of the entire disk. In this section, we set $a_{eq} = 10^{-5}$ cm and $\alpha = 10^{-4}$ (= constant), following the conclusion obtained in Sec. 3.1–3.2.

We assess the validities of $\Sigma_{inj}$ and $(\varepsilon^{54}Cr)_{PSD}$ using the proxy that the formation region of volatile-rich asteroids (i.e., CC parent bodies) would be estimated by bulk $\varepsilon^{54}Cr$ values of CCs (Table 1). The parent bodies of NCs and CCs were continuously accreted after CAI formation ($t = 0$). NCs were presumably formed in the reservoir separated from the CCs and earlier than the outer solar system. After the formation of the early formed planetesimals, the inward-drifting dust from the CC formation region can modify the isotopic compositions of NC parent bodies (Schiller et al. 2018). In addition, the secondary dust by the collision of planetesimals in the NC formation region also possibly modifies the compositions of individual NC parent bodies (Spitzer et al. 2020). For these reasons, it is difficult to interpret that the $\varepsilon^{54}Cr$ values of the inner solar system directly reflect the formation region of NCs. Nevertheless, given that CC parent bodies were accumulated immediately without any transportations of materials, we can estimate the relative formation region of parent bodies of CCs. We obtained age-corrected data by the individual accretional ages obtained by $^{26}Al$ heat-source variation (Sugiura & Fujiya 2014). Furthermore, we used the analytical errors of $\varepsilon^{54}Cr$ measurements in mass spectrometry (Table 1).

We simulate $\Sigma_{inj} = 1.0 \times 10^4$ to $5.0 \times 10^4$ and fixed $(\varepsilon^{54}Cr)_{PSD} = 2.5 \times 10^4$. First, we note that any condition with $\alpha = 10^{-3}$ does not reproduce four types of CCs (CV, CM, Tagish Lake, and CI) with a range of $r = 1$ to 20 au. The turbulence in this case was too strong to retain the $\varepsilon^{54}Cr$ heterogeneities observed in CCs (Sec. 3.1). In contrast, with $\alpha = 10^{-4}$, the formation regions of the four types of CCs were filled within a reasonable range of the disk radius (Fig. 8). The relative formation region of CCs moves to the inner parts as $\Sigma_{inj}$ values increase. Equations (9) and (20) suggest that the $\varepsilon^{54}Cr$ profiles varied not only with $\Sigma_{inj}$, but also with the $(\varepsilon^{54}Cr)_{PSD}$. With the same $\Sigma_{inj} \times (\varepsilon^{54}Cr)_{PSD}$ value, identical $\varepsilon^{54}Cr$ profiles were obtained.

The plausible values of $\Sigma_{inj}$ were evaluated based on the model of Ouellette et al. (2010), who proposed the "aerogel" model of the direct supernova injection of dust grains into the solar nebula. They showed that the supernova ejecta was clumpy. Moreover, the dust mass of each clump was approximately $2 \times 10^{-3}$ $M_{Sun}$ and the clump radius $r_{clump} \approx (1/300) \times d$, where $d$ is the distance between the supernova and the solar nebula. We can then roughly estimate the value of $\Sigma_{inj}$ as a function of $d$:

$$\Sigma_{inj} \sim \frac{M_{clump}}{\pi r_{clump}{}^2} \sim 1.2 \times 10^{-3} \times \left(\frac{d}{1\,pc}\right)^{-2} g\,cm^{-2}. \tag{21}$$

As the injection velocity is approximately $10^3$ km s$^{-1}$, the effect of gravitational focusing may be negligible. Ouellette et al. (2010) estimated that the distance of the supernovae injected into the solar system was approximately 2 pc ($\Sigma_{inj} \approx 3.0 \times 10^{-4}$ g cm$^{-2}$) from the abundances of short-lived nuclides recorded in meteorites. This value is consistent with the initial radius of the star cluster in which the Sun was born (approximately 1−3 pc; Portegies Zwart 2009). In the simulation with $(\varepsilon^{54}Cr)_{PSD} = 2.5 \times 10^4$ and $\Sigma_{inj} = (2-3) \times 10^{-4}$ g cm$^{-2}$ ($d = 2.0$ to 2.4 pc), four $r_{CC}$ (formation region of CCs) were obtained within a reasonable range (3 to 20 au; Fig. 8b–c). The simulation with $\Sigma_{inj} = 5.0 \times 10^{-4}$ g cm$^{-2}$ ($d = 1.5$ pc) was unrealistic because all $r_{CC}$ values were in the range of the gap formed by Jupiter (Fig. 8d). Recently, higher values of $(\varepsilon^{54}Cr)_{PSD}$ up to $5.0 \times 10^5$ were reported by Nittler et al. (2018). Assuming all the grains possessed these higher values, a reasonable range of $\Sigma_{inj}$ decreased to approximately $1.0 \times 10^{-5}$ g cm$^{-2}$, equivalent to $d \approx 10$ pc.

Desch et al. (2018) previously calculated the formation region of CC parent bodies based on the abundances of refractory materials (e.g., calcium–aluminum-rich inclusions) in individual CCs. The relative formation radius of the CC parent bodies obtained by Desch et al. (2018), namely, $r_{CV} < r_{CM} < r_{CI}$ was consistent with our model ($r_K < r_{Ch} < r_D < r_B$). The formation region of Tagish Lake was not determined in Desch et al. (2018), but the modal abundances of refractory materials in Tagish Lake can be calculated using major element components (Alexander 2019). The calculated abundance is approximately 0.6%, which is an intermediate of CI (B-type) and CM (Ch-type). Even if the estimation of Tagish Lake was included, the relative formation radius was fully consistent with Desch et al. (2018).

### 3.4 Dependence of the initial population of presolar dust

We demonstrated that $\varepsilon^{54}Cr$ heterogeneities were possibly explained by the mixing of solar and presolar dust with reasonable ranges of several parameters, including turbulent parameter ($\alpha$), dust size ($a_{eq}$), surface densities of supernovae injection ($\Sigma_{inj}$), and ($\varepsilon^{54}Cr$)$_{PSD}$ through Sec. 3.1 to 3.3. In addition, we infer the validity of the initial population of presolar dust to further certify the robustness of this model.

The mechanism responsible for the isotopic heterogeneities in the protoplanetary disk, including Cr, has been extensively discussed in previous studies. The late injection from a supernova is a possible candidate to explain the isotopic heterogeneities of meteorites. Here, we consider that supernova dust grains are injected into the surface of the solar nebula, in which case, the injected materials would be ubiquitously distributed within the disk radius (Ouellette et al. 2007). Meanwhile, the alternative initial distribution of the supernova injection can be considered as a narrow range injection into a certain area of the disk. Given that the supernova materials were injected into a certain area in the outer region, supernova dust grains would diffuse inward, which could reproduce the Cr isotopic evolution of bulk meteorites with temporal evolution (Sugiura & Fujiya 2014). In the narrow range injection model with a large $\alpha$, the earliest formed material (CAI) must possess the lowest $\varepsilon^{54}Cr$ values among the solar system. However, in reality, the $\varepsilon^{54}Cr$ of CAIs is the highest (approximately 6$\varepsilon$) among those materials.

The discrepancy about $\varepsilon^{54}Cr$ of CAIs can be solved in the context of the uniform supernovae injection assumed in this study. Previous studies proposed that the fine-grained presolar dust ($a_{eq} < 10^{-4}$ cm) incorporated into the CAI formation region would be evaporated under a high (>1600 K) disk temperature and gas phases would attain higher $\varepsilon^{54}Cr$ values than the solid phases remaining in the disk (Trinquier et al. 2009; Paton et al. 2013). The uniform injection to the disk can achieve a high abundance of presolar grains throughout the disk at $t = 0$, including the CAI formation region (<0.1 au). Our model also suggests that the inner solar system retains the $\varepsilon^{54}Cr$ heterogeneities ranging from –1 to 0, consistent with the $\varepsilon^{54}Cr$ compositions NCs (Trinquier et al. 2007).

Our results showed that $\varepsilon^{54}Cr$ variations predominantly originated from the retention of the spatial heterogeneity of the disk. The comparison between $\varepsilon^{54}Cr$ values and chronometric data also supports that spatial $\varepsilon^{54}Cr$ heterogeneities were preserved until CC formation. A study regarding Al-Mg dating (Nagashima et al. 2017) suggested that chondrules in CV chondrites formed 1 Myr before the accretion of parent bodies. However, the mean $\varepsilon^{54}Cr$ data of CV chondrules are indistinguishable from those of bulk CV chondrites (Olsen et al. 2016). This means that the bulk $\varepsilon^{54}Cr$ values for the CV formation region did not evolve effectively during 1 Myr, which is consistent with the $\varepsilon^{54}$Cr profiles obtained with $\alpha = 10^{-4}$ in our model.

### 3.5 Radial mixing of volatile-rich asteroids

The formation models of terrestrial planets were proposed with the constraint of several physical properties in the asteroid belt, such as mass, orbital eccentricities, and inclinations. One of the most important constraints in the asteroid belt is the radial mixing of compositionally different asteroids during the evolution of giant planets (Walsh et al. 2011; Raymond & Izidoro 2017; Clement et al. 2019). However, this constraint is generally difficult to use because of the unknown initial formation region of asteroids. We will evaluate a comprehensive evolutionary scenario using the initial asteroid locations obtained in this study and present asteroid distribution, generally following the concept of the Grand-Tack model (Walsh et al. 2011).

As discussed, the relative formation region of CCs can be estimated from our $\varepsilon^{54}$Cr isotopic heterogeneity model. The results for $\alpha = 10^{-4}$ and relatively smaller dust grains ($a_{eq} = 10^{-4}$ and $10^{-5}$ cm) preserve the isotopic heterogeneities until $t = 4$ Myr. For $\Sigma_{inj} = 3.0 \times 10^{-4}$ g cm$^{-2}$, consistent with the estimation from short-lived nuclides, a reasonable formation region of CCs was obtained (Fig. 8c).

The representative location of current asteroids may be preserved as the orbits of asteroid families. B-type asteroids dominantly existed in the Veritas family, located in 3.0–3.2 au. D-type asteroids were located further, 5.2 au, as Eurybates families. There is a discrepancy between the relative orbits of the B- and D-type asteroids between the initial ($r_D < r_B$ obtained in this study) and present ($r_B < r_D$) conditions (Fig. 9).

Importantly, the formation regions of Jupiter and Saturn have not been constrained. To explain the observed discrepancy in B- and D-type asteroids, Saturn may have been located between these types of asteroids. In this case, the orbits of the parent bodies for each asteroid type (i.e., B, Ch, D, and K types) were disturbed when the migration of giant planets started. First, Jupiter migrated inward, which did not significantly affect the orbits of the four asteroid types but caused the scatter of inclination and orbits of S-type asteroids. A subsequent inward migration of Saturn induced scattering of the K-, Ch-, and D-type asteroids both inwardly and outwardly. Remarkably, a part of the D-type asteroids may be transported beyond the formation region of the B-type asteroids. Finally, the outward migration of Jupiter and Saturn mainly transported K-, Ch-, and B-type asteroids to the main belt. The formation regions of the K-, Ch-, and B-type asteroids were reflected in the compositional zoning of the main belt. D-type asteroids were possibly delivered to the Trojan regions at approximately 3.6–3.8 Gyr by late heavy bombardment from trans-Neptune (Morbidelli et al. 2005; Levison et al. 2009). This scenario does not include the dynamic properties of asteroids, such as inclinations and eccentricities, instead, we emphasized the significance of the scenario proposed herein as the first case of linking the isotopic records of meteorites and the population of asteroids.

# Conclusion

In this study, we conducted numerical calculations for the viscous evolution of gas and isotopically different (solar and presolar) dust in the protoplanetary disk. We assumed that the presolar grains from the supernovae were uniformly injected into the surface of the disk. From our assumption, it is self-evident that the outer solar system shows a high relative concentration of presolar grains. The preservation of the isotopic heterogeneity is dominantly associated with the turbulent parameter, $\alpha$. With a relatively high turbulence parameter ($\alpha = 10^{-3}$), the disk was quickly homogenized and the Cr isotopic variations of CCs could not be reproduced. We found that the isotopic heterogeneity was well preserved with a low turbulent parameter ($\alpha = 10^{-4}$). In this simulation, the isotopic homogenization of the outer solar system was prevented with a lower dust size ($a_{eq} = 10^{-4}-10^{-5}$ cm).

The gap formed by giant planets is considered to be a cause of the Cr isotopic dichotomy between NCs and CCs. We tested two treatments of turbulence in the vicinity of the gap; (i) $\alpha_{diff} = \alpha_{acc}$ and (ii) $\alpha_{diff} =$ constant. In the case of $\alpha_{diff} =$ const., clear compositional gaps were reproduced between the NC and CC formation regions.

With these settings (low $\alpha$, low dust size, and gap), suitable $\Sigma_{inj}$ values (surface densities of the injected materials) were investigated. We estimated a reasonable formation region of the four types of CCs using the profiles of $\varepsilon^{54}$Cr with varied disk radii. With $\Sigma_{inj} = 3 \times 10^{-4}$ g cm$^{-2}$ equivalent to 2 pc of the distance from supernovae, the formation region of four types of CCs can be reproduced in a reasonable disk radius (3–10 au). This is consistent with the abundances from short-lived nuclides (Ouellette et al. 2010).

Among the volatile-rich asteroids associated with investigated CCs, the obtained relative orbits from these asteroids are $r_K < r_{Ch} < r_D < r_B$. This sequence is consistent with the individual model of Desch et al. (2018). However, these results are inconsistent with the present orbits implied from the orbits of their asteroid families ($r_K < r_{Ch} < r_B < r_D$). We speculate that the orbital changes were caused by the migration of Jupiter and Saturn. Remarkably, the final outward migration of Jupiter and Saturn induced inward transportation of the K-, Ch-, and B-type asteroids to the main belt, which explains the discrepancy between the initial and current locations.

We thank to R. Hyodo and S. Mori for their fruitful discussion. We are grateful to Daniel J. Scheeres and an anonymous reviewer for the editorial handling and their constructive reviews. This research was supported by Grants-in-Aid for Scientific Research from Japan Society for the Promotion of Science (JP18J14186, JP20J00598, and JP20K14535).

# Tables

Table 1: Compilation of Cr Isotopic Compositions of carbonaceous chondrites (CCs)

| Meteorites | Subgroups | $\varepsilon^{54}Cr$ | Errors | References |
|------------|-----------|----------------------|--------|------------|
| Allende | CV | 0.83 | 0.05 | (1)–(4) |
| Murchison | CM | 0.98 | 0.04 | (2)–(3), (5) |
| Tagish Lake | C2-ung. | 1.19 | 0.15 | (6) |
| Orgueil | CI | 1.58 | 0.03 | (1)–(3), (6) |

References: (1) Shukolyukov & Lugmair (2006), (2) Trinquier et al. (2007) (3) Qin et al. (2010), (4) Hibiya et al. (2018), (5) Jenniskens et al. (2012), and (6) Petitat et al. (2011)

# Appendix

*Low-mass proto-Jupiter*

We discussed the effects of the early formed Jupiter ($20M_E$) on the $\varepsilon^{54}Cr$ profiles in the disk. We also tested the case of proto-Jupiter ($3M_E$) located in 3.5 au, as suggested by the low-mass asteroid model (Raymond & Izidoro 2017). Figs. A1–A2 show the results of the $\varepsilon^{54}Cr$ distribution and surface densities of the dust. The $\varepsilon^{54}Cr$ profiles show no changes from the simulation without giant planets in Sec. 3.1. The surface densities of the dust show smaller depletion around 3.5 au. The compositional gap in $\varepsilon^{54}Cr$ was not reproduced in any $\alpha$ or $a_{eq}$.

*Formation of Saturn*

In contrast to Jupiter, the formation timing and location of Saturn are not well constrained. Given that Saturn formed at $t = 0.5$ Myr as Jupiter, the additional gap opens in the orbit beyond Jupiter. We tested the case of proto-Jupiter ($20M_E$) located in 3.5 au and proto-Saturn ($20M_E$) located in 4.5 au. This combination was suggested by (Walsh, et al. 2011). Figs. A3–A4 show the results of the $\varepsilon^{54}Cr$ distribution and surface densities of the dust. The $\varepsilon^{54}Cr$ profiles show no changes from the simulation without proto-Saturn. Fig. A4 suggests that a gap opened by Saturn combined with that of Jupiter. This also suggests that most CCs should have formed beyond Saturn in this situation.

We also tested the case of proto-Jupiter ($20M_E$) located in 3.5 au and proto-Saturn ($20M_E$) located in 10 au. Figs. A5–A6 show the results of the $\varepsilon^{54}Cr$ distribution and surface densities of the dust. With $\alpha = 10^{-4}$, two distinct gaps were observed in the profiles of $\varepsilon^{54}Cr$. This indicates that the $\varepsilon^{54}Cr$ values of almost all CCs were filled with the gap range formed by Saturn.

# Figure captions

Fig. 1: $\varepsilon^{54}$Cr evolution with the disk radius at $t = 0$ Myr, 2 Myr, and 4 Myr. The existence of giant planets is not assumed.

Fig. 2: Timescales of advection (dotted lines), diffusion (dashed lines), and radial drift (bold lines) with the disk radius.

Fig. 3: Surface density evolution of solar dust with the disk radius at $t = 0$ Myr, 2 Myr, and 4 Myr. The existence of giant planets was not assumed.

Fig. 4: Stokes numbers with the disk radius.

Fig. 5: $\varepsilon^{54}$Cr evolution with the disk radius at $t = 0$ Myr, 2 Myr, and 4 Myr. The existence of Jupiter ($20M_E$) is assumed at $r = 3$ au. Here, $\alpha_{\mathrm{diff}}$ was assumed to be identical to $\alpha_{\mathrm{acc}}$.

Fig. 6: Surface density evolution of solar dust with the disk radius at $t = 0$ Myr, 2 Myr, and 4 Myr. The existence of Jupiter ($20M_E$) is assumed at $r = 3$ au.

Fig. 7: $\varepsilon^{54}$Cr evolution with the disk radius at $t = 0$ Myr, 2 Myr, and 4 Myr. The existence of Jupiter ($20M_E$) is assumed at $r = 3$ au. Here, $\alpha_{\mathrm{diff}}$ was assumed to be constant ($10^{-4}$).

Fig. 8: Estimated relative formation location with accretional ages of K- (red), Ch- (green), D- (blue), and B- (black) type asteroids. Dotted lines show the analytical uncertainties of $\varepsilon^{54}$Cr.

Fig. 9: Schematic image of the asteroid's evolution with the migration of giant planets.

Fig. A1: $\varepsilon^{54}$Cr evolution with the disk radius at $t = 0$ Myr, 2 Myr, and 4 Myr. The existence of Jupiter ($3M_E$) is assumed at $r = 3$ au.

Fig. A2: Surface density evolution of solar dust with the disk radius at $t$ = 0 Myr, 2 Myr, and 4 Myr. The existence of Jupiter ($3M_E$) is assumed at $r$ = 3 au.

Fig. A3: $\varepsilon^{54}Cr$ evolution with the disk radius at $t$ = 0 Myr, 2 Myr, and 4 Myr. The existences of Jupiter ($20M_E$) and Saturn ($20M_E$) are assumed at $r$ = 3 and 4.5 au, respectively.

Fig. A4: Surface density evolution of solar dust with the disk radius at $t$ = 0 Myr, 2 Myr, and 4 Myr. The existences of Jupiter ($20M_E$) and Saturn ($20M_E$) are assumed at $r$ = 3 and 4.5 au, respectively.

Fig. A5: $\varepsilon^{54}Cr$ evolution with the disk radius at $t$ = 0 Myr, 2 Myr, and 4 Myr. The existences of Jupiter ($20M_E$) and Saturn ($20M_E$) are assumed at $r$ = 3 and 10 au, respectively.

Fig. A6: Surface density evolution of solar dust with the disk radius at $t$ = 0 Myr, 2 Myr, and 4 Myr. The existence of Jupiter ($20M_E$) and Saturn ($20M_E$) are assumed at $r$ = 3 and 10 au, respectively.

# References

Alexander, C. M. O. D. 2019, GeCoA, 254, 277

Amari, S., Lewis, R. S., & Anders, E. 1994, GeCoA, 58, 459

Burbine, T. H. 1998, M&PS, 33, 253

Burbine, T. H., Binzel, R. P., Bus, S., & Clark, B. E. 2001, M&PS, 36, 245

Chiang, E. I., & Goldreich, P. 1997, ApJ, 490, 368

Ciesla, F. J. 2010, Icar, 208, 455

Clement, M. S., Raymond, S. N., & Kaib, N. A. 2019, ApJ, 157

Cloutis, E. A., Hiroi, T., Gaffey, M. J., et al. 2011a, Icar, 212, 180

Cloutis, E. A., Hudon, P., Hiroi, T., & Gaffey, M. J. 2012, Icar, 217, 389

Cloutis, E. A., Hudon, P., Hiroi, T., et al. 2011b, Icar, 216, 309

Dauphas, N., Remusat, L., Chen, J. H., et al. 2010, ApJ, 720, 1577

Dauphas, N., & Schauble, E. A. 2016, AREPS, 44, 709

Desch, S. J., Estrada, P. R., Kalyaan, A., & Cuzzi, J. N. 2017, ApJ, 840, 86

Desch, S. J., Kalyaan, A., & Alexander, C. M. O. D. 2018, ApJS, 238, 11

Dullemond, C. P., & Dominik, C. 2005, A&A, 434, 971

Fukai, R., & Yokoyama, T. 2017, E&PSL, 474, 206

Fukai, R., & Yokoyama, T. 2019, ApJ, 879, 79

Greenwood, R. C., Burbine, T. H., & Franchi, I. A. 2020, GeCoA, 277, 377

Hartmann, L., Calvet, N., Gullbring, E., & D'Alessio, P. 1998, ApJ, 495, 385

Haugbølle, T., Weber, P., Wielandt, D. P., et al. 2019, AJ, 158

Hayashi, C. 1981, PThPS, 70, 35

Hibiya, Y., Iizuka, T., Yamashita, K., et al. 2018, G&GR, 43, 133

Hiroi, T., Zolensky, M., & Pieters, C. M. 2001, Sci, 293, 2234

Homma, K., & Nakamoto, T. 2018, ApJ, 868, 118

Hoppe, P., Leitner, J., & Kodolányi, J. 2017, Natur Astron, 1, 617

Jacquet, E., Pignatale, F. C., Chaussidon, M., & Charnoz, S. 2019, ApJ, 884, 32

Jenniskens, P., Fries D. Marc, Yin, Q. Z., M., Z., et al. 2012, Sci, 338, 1583

Kanagawa, K. D., Muto, T., Tanaka, H., et al. 2015a, ApJL, 806. 15

Kanagawa, K. D., Tanaka, H., Muto, T., et al. 2015b, MNRAS, 448, 994

Kanagawa, K. D., Muto, T., Tanaka, H., et al. 2016, PASJ, 68, 43

Kozasa, T., Hasegawa, H., & Nomoto, K. 1991, A&A, 249, 474

Kruijer, T. S., Burkhardt, C., Budde, G., & Kleine, T. 2017, PNAS, 114, 6712

Kruijer, T. S., Kleine, T., & Borg, L. E. 2019, Natur Astron, 4, 32

Kusaka, T., Nakano, T., & Hayashi, C. 1970, PThP, 44, 1580

Larsen, K. K., Trinquier, A., Paton, C., et al. 2011, ApJL, 735, 37

Levison, H. F., Bottke, W. F., Gounelle, M., et al. 2009, Natur, 460, 364

Lodders, K. 2003, ApJ, 591, 1220

Morbidelli, A., Chambers, J. E., Lunine, J. I., et al. 2000, M&PS, 35, 1309

Morbidelli, A., Levison, H. F., Tsiganis, K., & Gomes, R. 2005, Natur, 435, 462

Mousis, O., Ronnet, T., & Lunine, J. I. 2019, ApJ, 875, 9

Nagashima, K., Krot, A. N., & Komatsu, M. 2017, GeCoA, 201, 303

Nanne, J. A. M., Nimmo, F., Cuzzi, J. N., & Kleine, T. 2019, E&PSL, 511, 44

Nittler, L. R., CM, O. D. A., Liu, N., & Wang, J. 2018, ApJL, 856, 24

Okuzumi, S. 2009, ApJ, 698, 1122

Okuzumi, S., Tanaka, H., Kobayashi, H., & Wada, K. 2012, ApJ, 752, 106

Olsen, M. B., Wielandt, D., Schiller, M., et al. 2016, GeCoA, 191, 118

Ouellette, M., Desch, S. J., & Hester, J. J. 2007, ApJ, 662, 1268

Ouellette, N., Desch, S. J., Bizzarro, M., et al. 2009, GeCoA, 73, 4946

Ouellette, N., Desch, S. J., & Hester, J. J. 2010, ApJ, 711, 597

Paton, C., Schiller, M., & Bizzarro, M. 2013, ApJL, 763, 40

Petitat, M., Birck, J. L., Luu, T. H., & Gounelle, M. 2011, ApJ, 736, 23

Qin, L., Alexander, C. M. O. D., Carlson, R. W., et al. 2010, GeCoA, 74, 1122

Qin, L., Nittler, L. R., Alexander, C. M. O. D., et al. 2011, GeCoA, 75, 629

Raymond, S. N., & Izidoro, A. 2017, Icar, 297, 134

Schiller, M., Bizzarro, M., & Fernandes, V. A. 2018, Natur, 555, 507

Shakura, N. I., & Sunyaev, R. A. 1973, A&A, 24, 337

Shukolyukov, A., & Lugmair, G. 2006, E&PSL, 250, 200

Spitzer, F., Burkhardt, C., Budde, G., et al. 2020, ApJL, 898, 2

Sugiura, N., & Fujiya, W. 2014, M&PS, 49, 772

Trinquier, A., Birck, J.-L., & Allègre, C. J. 2007, ApJ, 655, 1179

Trinquier, A., Elliott, T., Ulfbeck, D., et al. 2009, Sci, 324, 374

Vernazza, P., Castillo-Rogez, J., Beck, P., et al. 2017, AJ, 153, 72

Vernazza, P., Marsset, M., Beck, P., et al. 2015, ApJ, 806, 204

Walsh, K. J., Morbidelli, A., Raymond, S. N., et al. 2011, Natur, 475, 206

Warren, P. H. 2011, E&PSL, 311, 93

Yang, L., & Ciesla, F. J. 2012, M&PS, 47, 99

Yokoyama, T., Nagai, Y., Fukai, R., & Hirata, T. 2019, ApJ, 883, 62

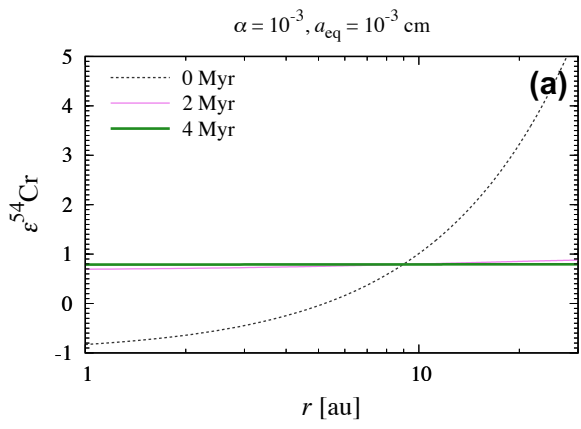
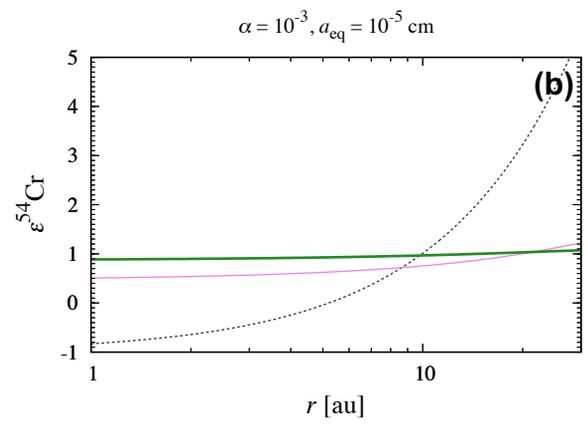
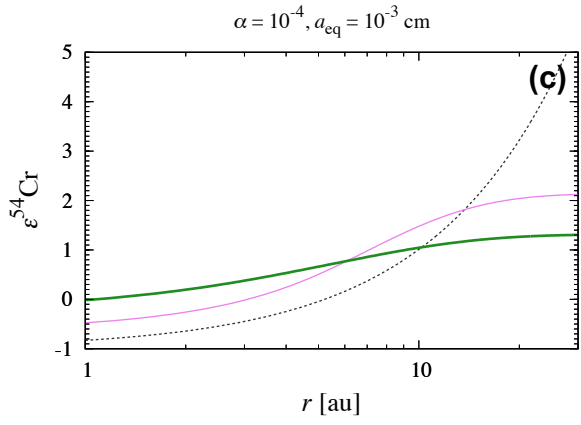
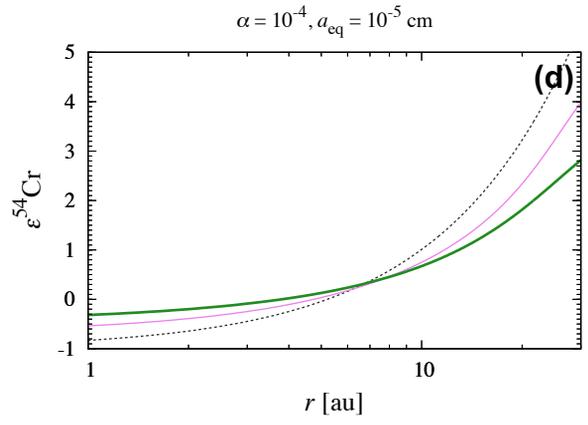

**Fig. 1**:

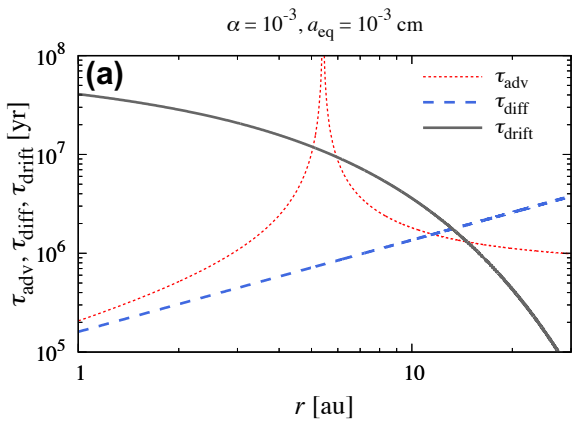
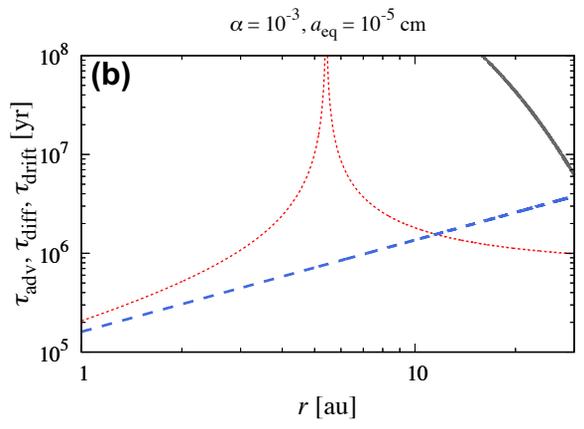
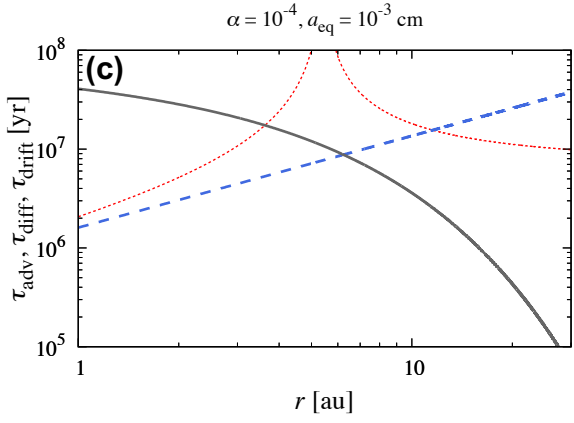
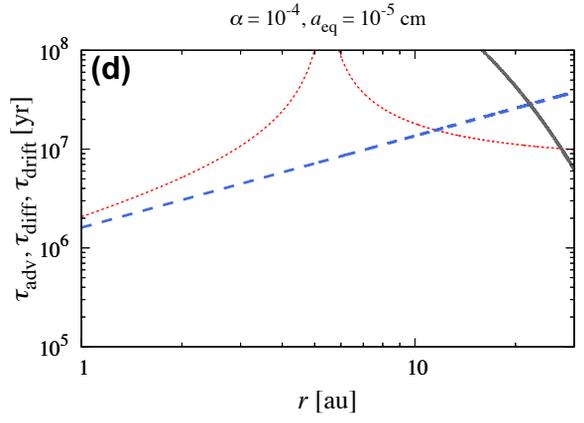

**Fig. 2:**

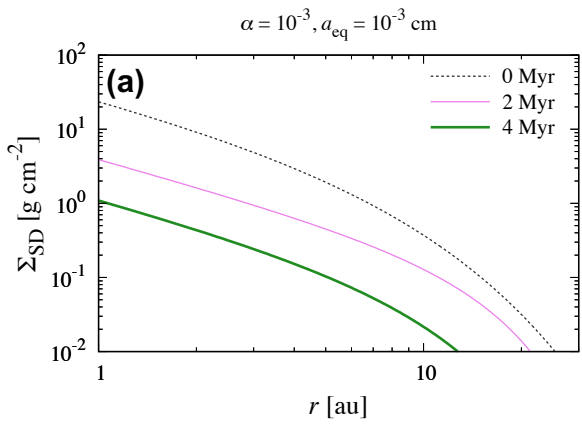
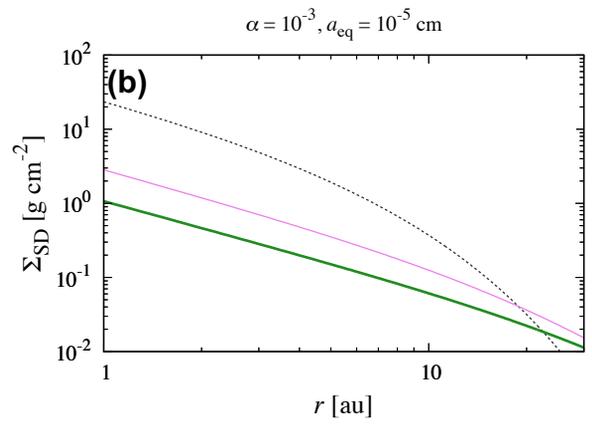
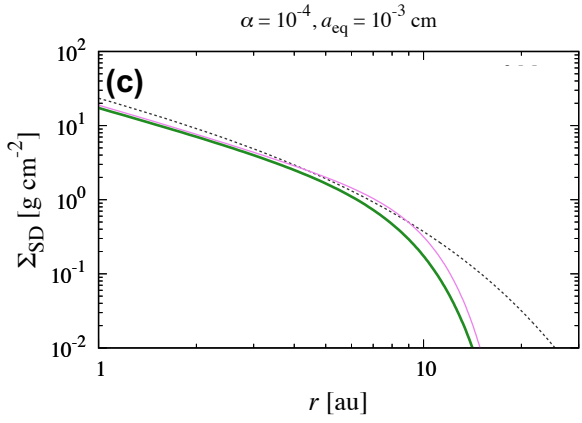
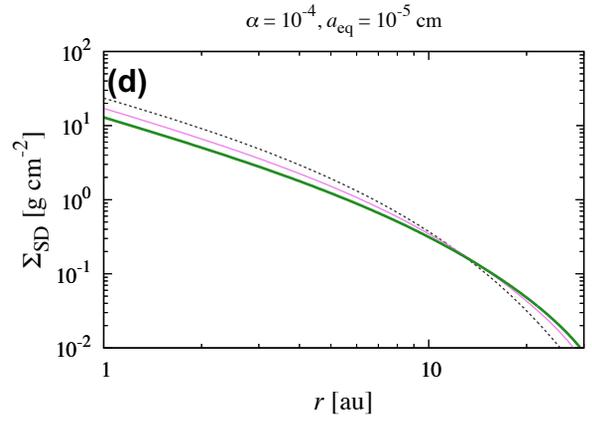

**Fig. 3**:

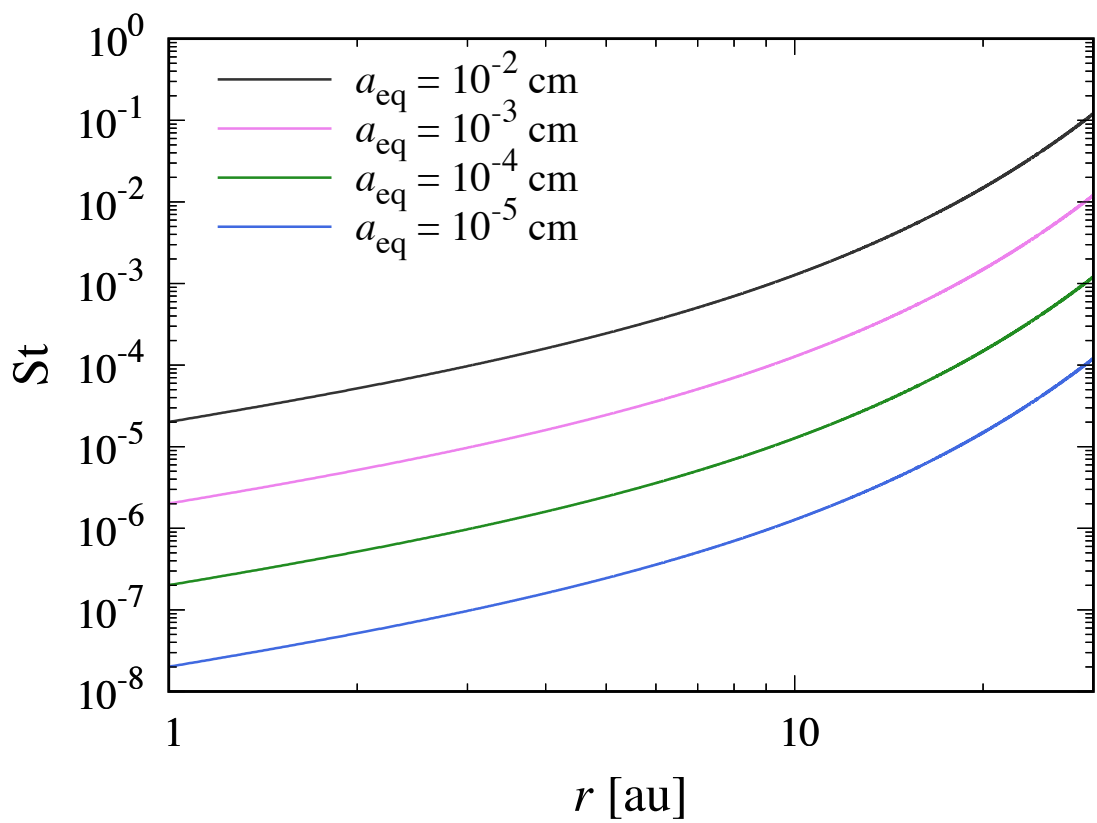

**Fig. 4**:

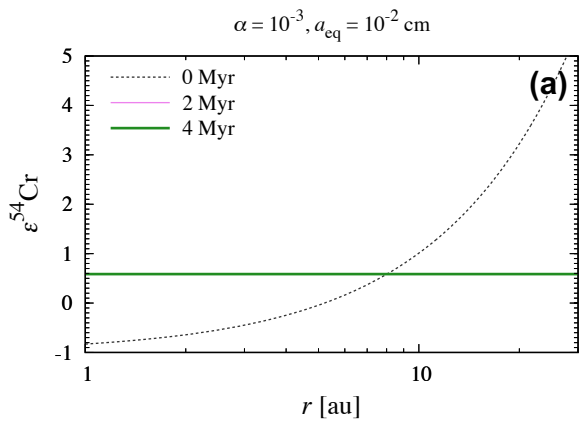
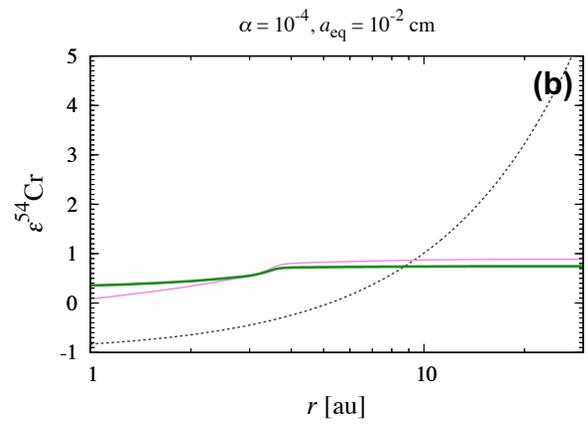

**Fig. 5**:

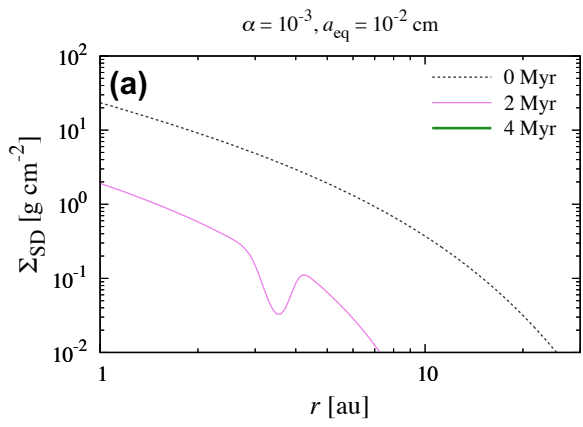
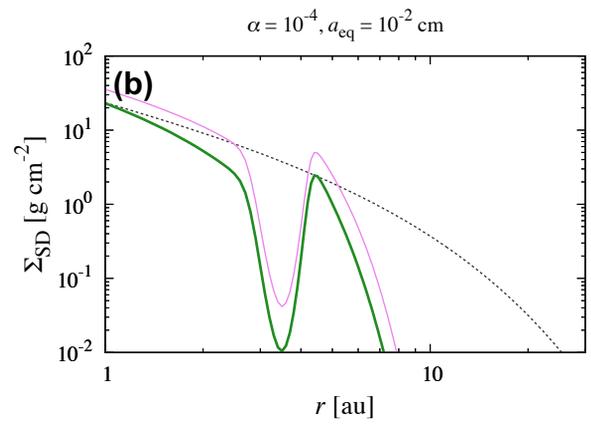

**Fig. 6**:

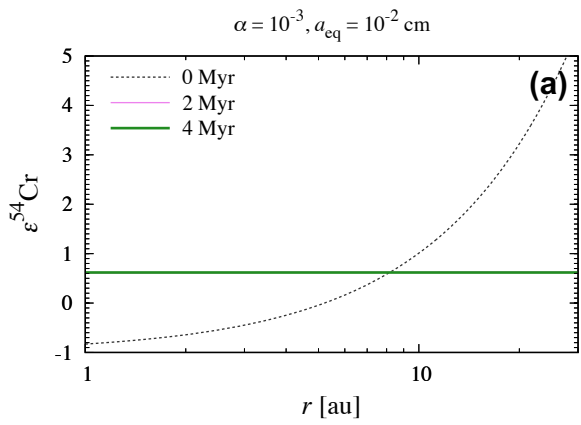

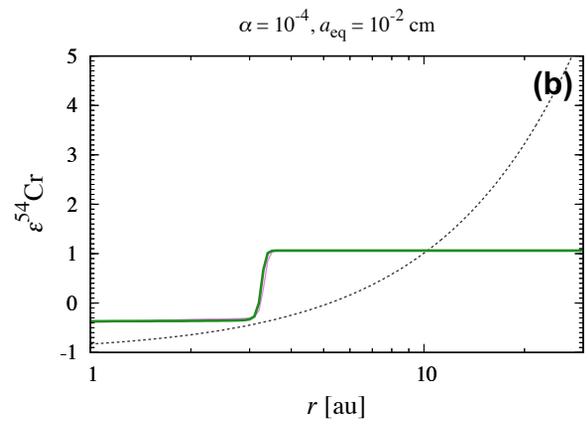

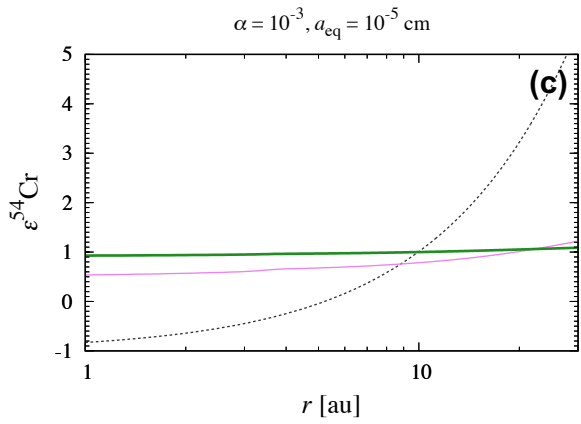

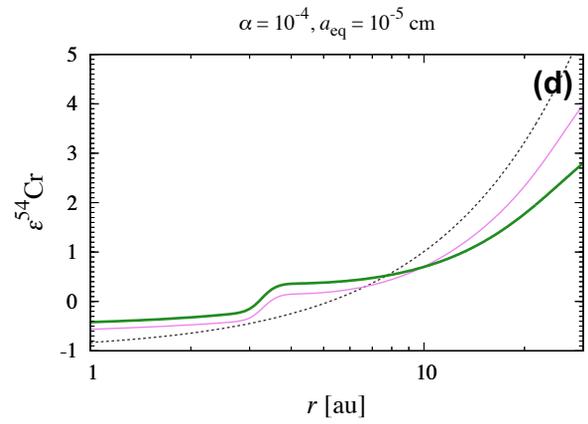

**Fig. 7:**

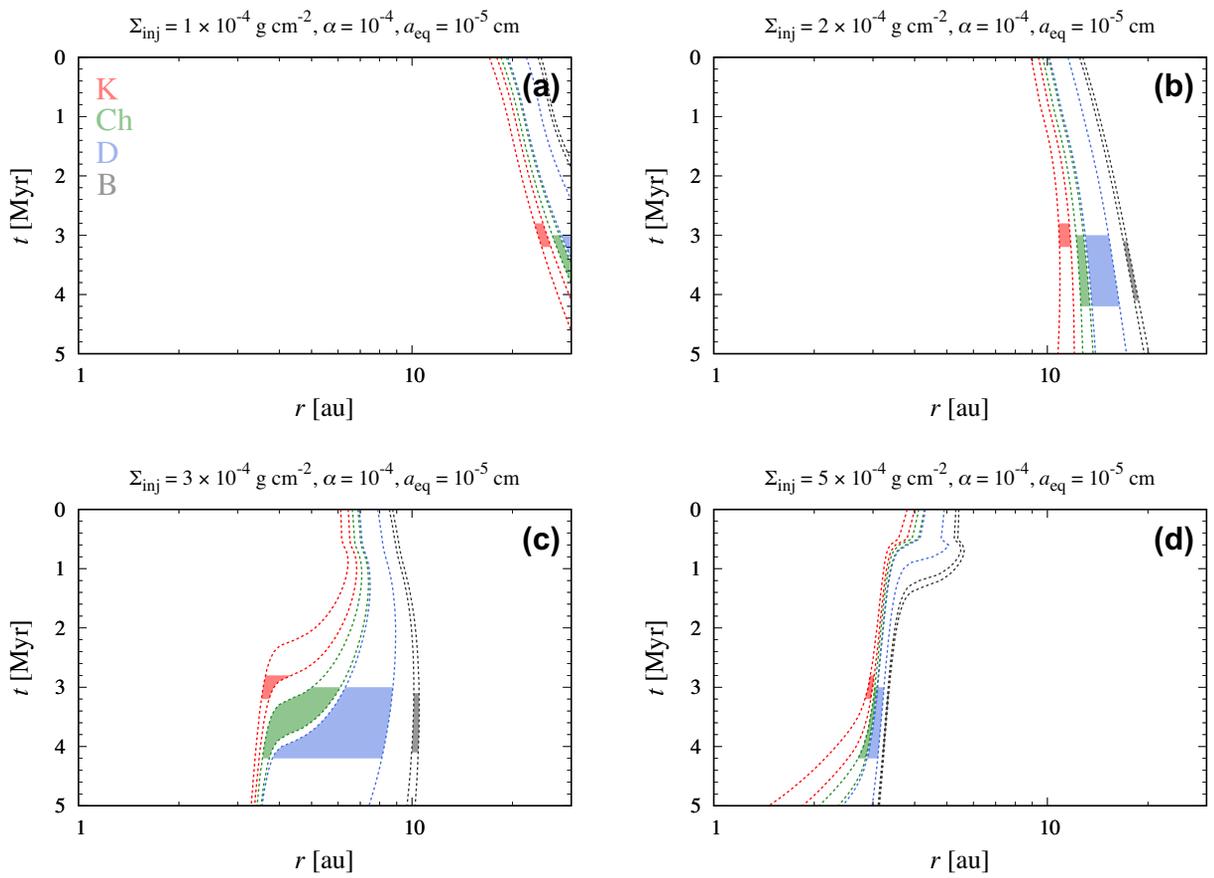

**Fig. 8:**

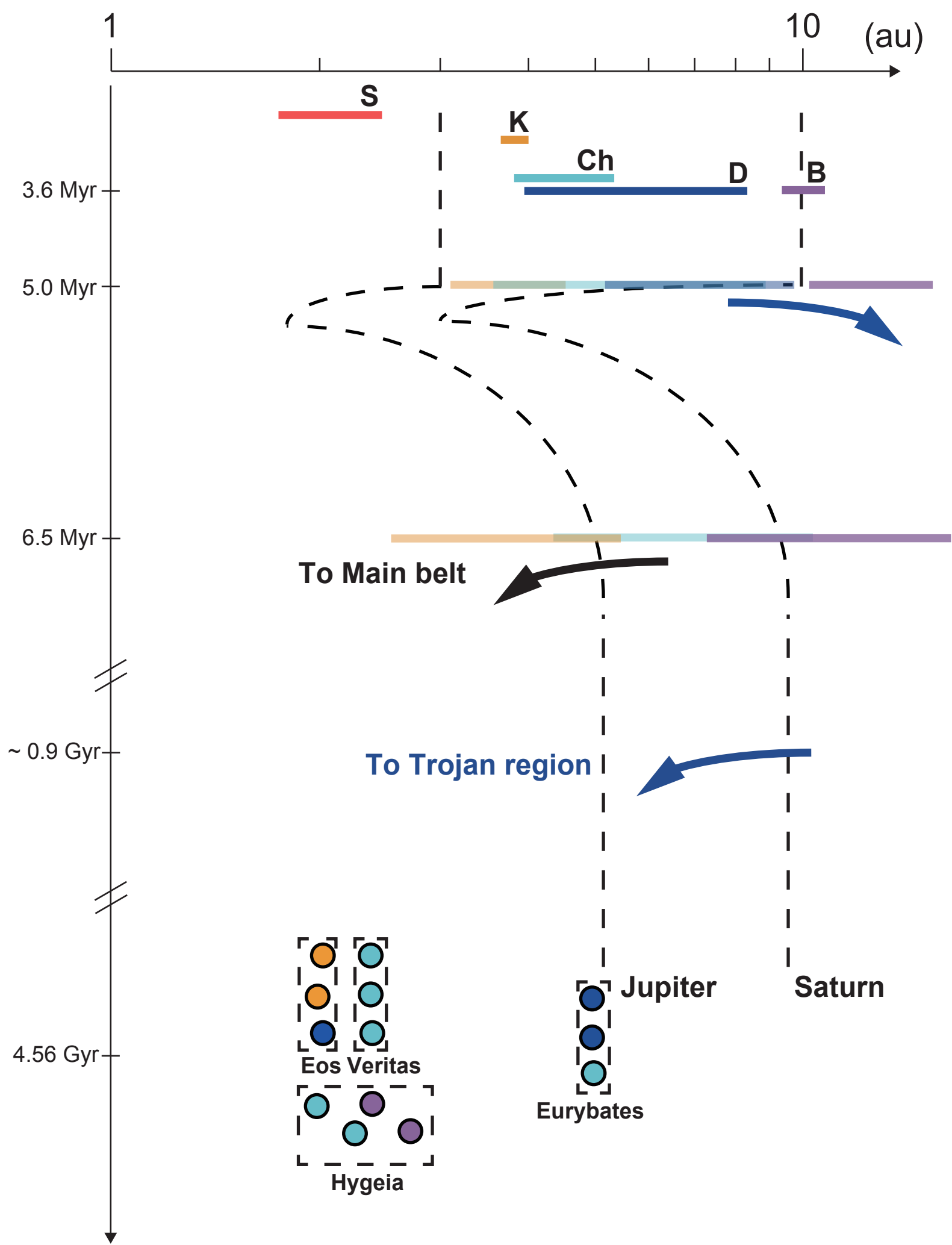

**Fig. 9:**

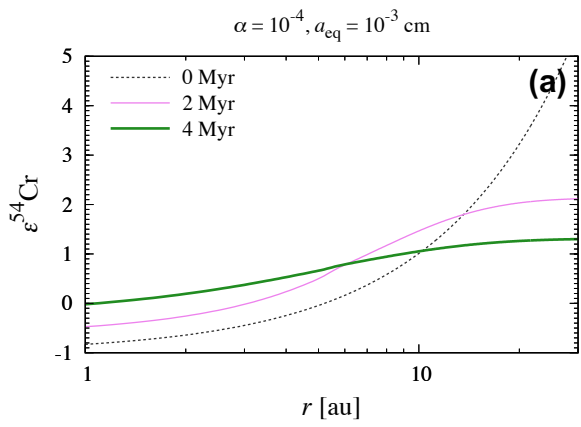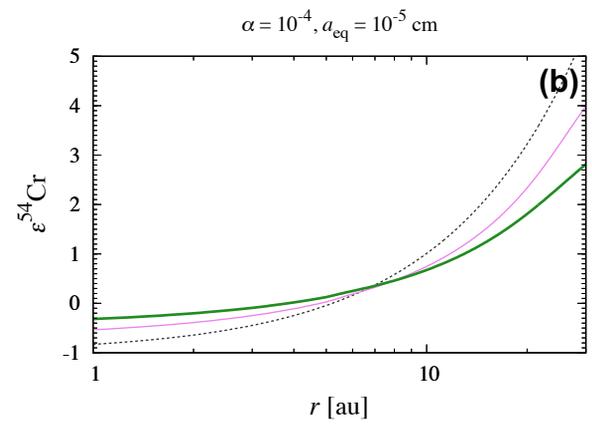

**Fig. A1**:

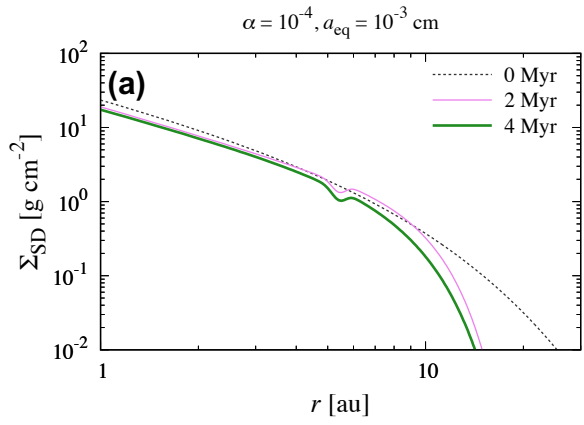
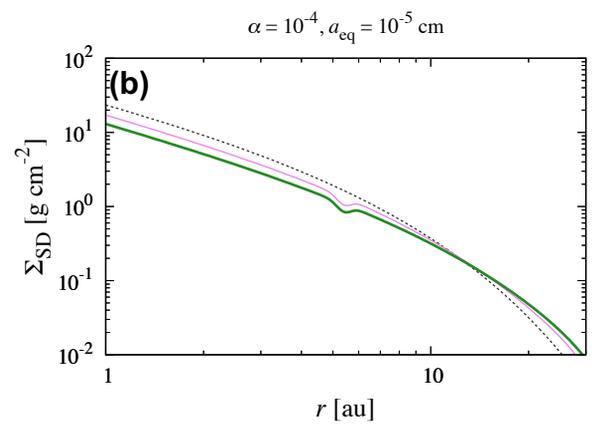

**Fig. A2**:

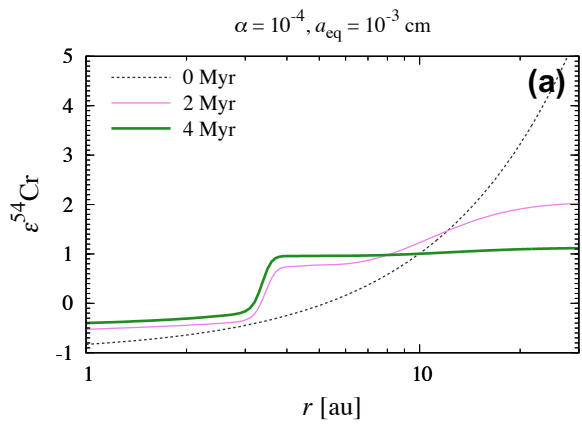
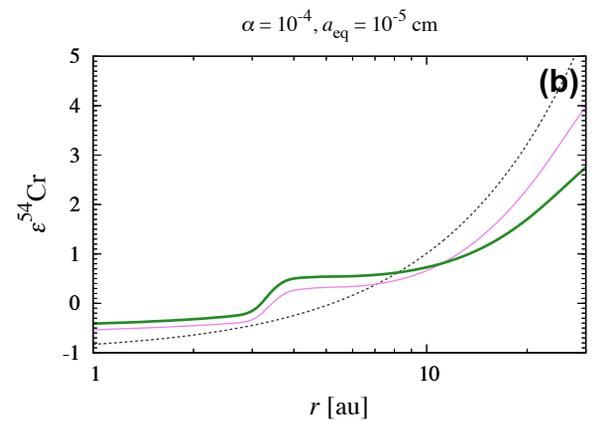

**Fig. A3**:

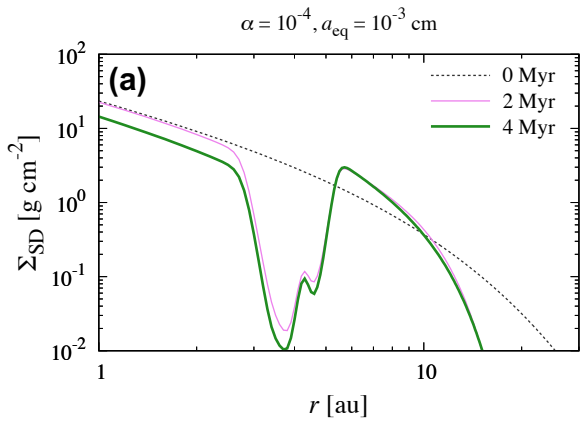
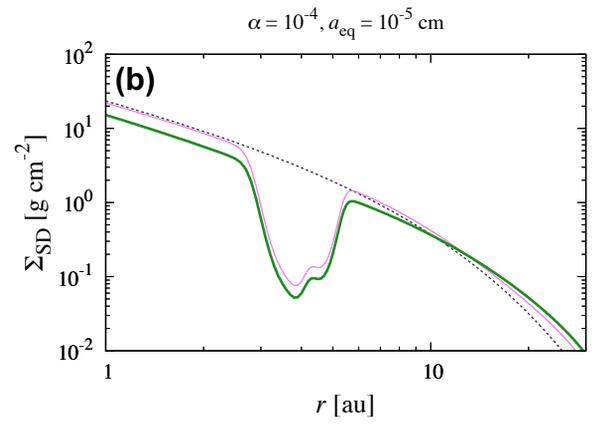

**Fig. A4**:

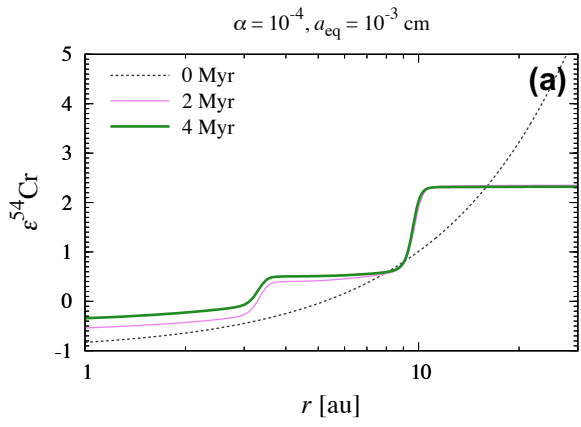
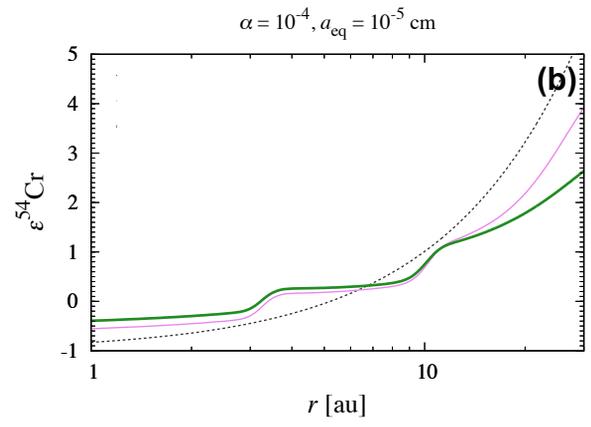

**Fig. A5**:

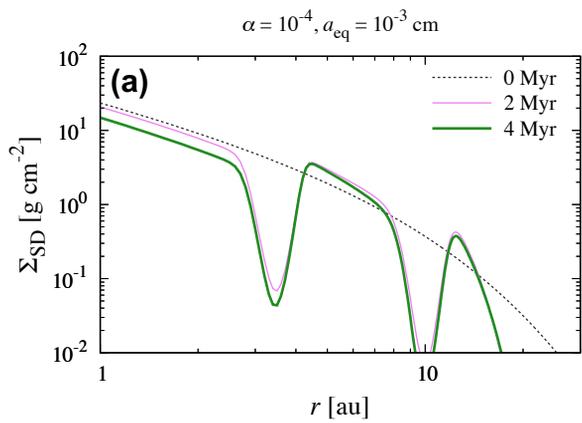

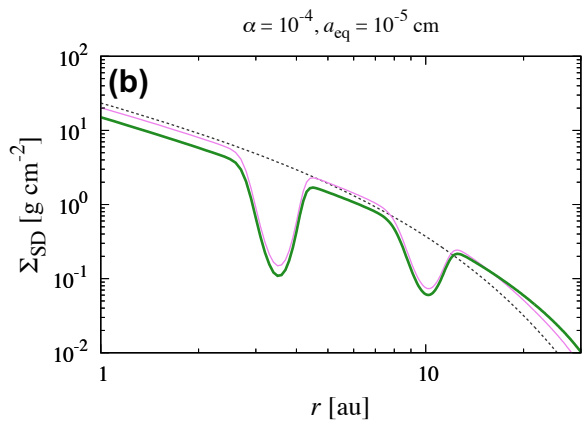

**Fig. A6**: